\documentclass{article}
\usepackage{spconf,amsmath,graphicx}

\usepackage{cite}
\usepackage{url}
\usepackage{bm}
\usepackage{hyperref}
\usepackage{xspace}
\usepackage{verbatim}
\usepackage{multirow}
\usepackage{booktabs}
\usepackage[table,xcdraw]{xcolor}
\usepackage{array}
\usepackage{amsmath,amssymb}
\usepackage{tabularx}
\newcolumntype{Y}{>{\centering\arraybackslash}X}

\usepackage{setspace}

\def\c{{\bm{c}}}
\def\x{{\bm{x}}}

\def\L{{\cal L}}

\newcommand\figref[1]{Fig.~\ref{#1}}
\newcommand\tabref[1]{Table~\ref{#1}}
\newcommand\secref[1]{Section~\ref{#1}}
\newcommand\mathref[1]{Equation~(\ref{#1})}

\title{Source-Filter HiFi-GAN: Fast and Pitch Controllable \\ High-Fidelity Neural Vocoder}
\name{Reo Yoneyama$^{1}$, Yi-Chiao Wu$^{2}$, and Tomoki Toda$^{1}$}
\address{$^{1}$Nagoya University, Japan, $^{2}$Meta Reality Labs Research, USA}

\begin{document}
\fontsize{9.77}{11.5}\selectfont
\maketitle
\begin{abstract}
Our previous work, the unified source-filter GAN (uSFGAN) vocoder, introduced a novel architecture based on the source-filter theory into the parallel waveform generative adversarial network to achieve high voice quality and pitch controllability. However, the high temporal resolution inputs result in high computation costs. Although the HiFi-GAN vocoder achieves fast high-fidelity voice generation thanks to the efficient upsampling-based generator architecture, the pitch controllability is severely limited. To realize a fast and pitch-controllable high-fidelity neural vocoder, we introduce the source-filter theory into HiFi-GAN by hierarchically conditioning the resonance filtering network on a well-estimated source excitation information. According to the experimental results, our proposed method outperforms HiFi-GAN and uSFGAN on a singing voice generation in voice quality and synthesis speed on a single CPU. Furthermore, unlike the uSFGAN vocoder, the proposed method can be easily adopted/integrated in real-time applications and end-to-end systems.
\end{abstract}


\section{Introduction}
\label{sec:intro}
A neural vocoder is a generative deep neural network (DNN) that generates raw waveforms based on the input acoustic features. 
The vocoder capability (e.g., voice quality, controllability, and synthesis speed) dramatically affects the overall performance of speech generation applications such as text-to-speech (TTS), singing voice synthesis (SVS), and voice conversion.
Especially, the controllability of fundamental frequencies ($\text{F}_0$) is essential for flexibly generating desired intonation and musical pitch patterns.

HiFi-GAN \cite{hifigan} is one of the most popular neural vocoders due to its convincing voice quality and efficient synthesis.
HiFi-GAN gradually upsamples the low temporal resolution acoustic features to match the high temporal resolution raw waveform. 
As the computational cost depends on the sequence length, the upsampling architecture, which works from much lower temporal resolutions, facilitates fast synthesis.
Although HiFi-GAN and recent high-fidelity neural vocoders \cite{style-melgan, hn-pwg, diffwave, wavegrad} have achieved convincing voice quality, they lack $\text{F}_0$ controllability.

To solve the aforementioned problem, unified source-filter GAN (uSFGAN) \cite{usfgan, hn-usfgan} attempts to introduce a reasonable inductive bias of human voice production to DNNs to simultaneously achieve high voice quality and $\text{F}_0$ controllability.
USFGAN factorizes the generator network of Quasi-Periodic Parallel WaveGAN \cite{qppwg} into a source excitation generation network (source-network) and a resonance filtering network (filter-network).
However, because of the very high temporal resolution inputs, the slow synthesis makes it challenging to adopt/integrate uSFGAN in real-time applications, and end-to-end systems \cite{vits, jets, visinger}.

For fast high-fidelity generations and $\text{F}_0$ controllability, we propose source-filter HiFi-GAN (SiFi-GAN), which introduces source-filter modeling into HiFi-GAN.
The proposed model is based on the V1 configuration described in the HiFi-GAN paper \cite{hifigan} because of its superior voice quality.
SiFi-GAN has two separate upsampling networks: a source-network and a filter-network, which are connected in series to simulate a pseudo cascade mechanism of the source-filter theory \cite{source_filter}.
The parameters of HiFi-GAN V1 are pruned to compensate for the additional computation costs of the source-filter modeling.
According to the experimental results, SiFi-GAN achieves a faster synthesis than HiFi-GAN V1 on a single CPU with better voice quality. Also, SiFi-GAN achieves comparable $\text{F}_0$ controllability as the uSFGAN-based model with much faster synthesis.

\section{Baseline HiFi-GAN and uSFGAN}
\label{ssec:related work}
This chapter describes two neural vocoders: HiFi-GAN \cite{hifigan} and uSFGAN \cite{usfgan, hn-usfgan}. 
They are the basis of the proposed SiFi-GAN and we use them as the baselines.

\subsection{HiFi-GAN}
HiFi-GAN \cite{hifigan} is a generative adversarial networks (GAN) \cite{gan} based high-fidelity neural vocoder with a sophisticatedly designed generator and multi-period and multi-scale discriminators.
The generator receives a mel-spectrogram as input and upsamples it to match the temporal resolution of the target raw waveform with transposed convolution neural networks followed by multi-receptive field fusion (MRF) modules.
MRF comprises several residual blocks, each has customized kernel and dilation sizes to capture the input feature from different receptive fields. 
The hyperparameters in MRFs can be defined as a trade-off between synthesis efficiency and voice quality.

The generator learns to fool the discriminators while the discriminators learn to distinguish between the generated and ground truth speech, following least square GAN \cite{lsgan}.
The training objective of the generator $\L_{G}$ comprises three losses: an adversarial loss $\L_{g, \text{adv}}$, a feature matching loss \cite{melgan} $\L_{\text{fm}}$, and a mel-spectral L1 loss $\L_{\text{mel}}$.
\begin{equation}
    \L_{G} = \L_{g, \text{adv}} + \lambda_{\text{fm}} \L_{\text{fm}} + \lambda_{\text{mel}}\L_{\text{mel}}
\end{equation}
where, $\lambda_{\text{fm}}$ and $\lambda_{\text{mel}}$ are loss balancing hyperparameters.

\subsection{Unified Source-Filter GAN}

\subsubsection{Source excitation regularization loss}
USFGAN \cite{usfgan, hn-usfgan} decomposes a single DNN into the source-network and the filter-network using a regularization loss on the output of the source-network. 
The regularization loss is designed to facilitate the source-network to generate signals with reasonable source characteristics by taking into account the physical limitations of the actual source excitation signals.
For instance, the loss metric adopted in \cite{hn-usfgan} approximates the output source excitation signal to the residual signal in linear predictive coding \cite{lpc, lpc_1982, lpc_1995}.
The loss is formulated as follows:
\begin{equation} \label{math:reg-loss}
    \L_{\text{reg}} = \mathbb{E}_{\x, \c} \left[ \frac{1}{N} \lVert ~ \log{\psi(S)} - \log{\psi(\hat{S})} ~ \rVert_{1} \right],
\end{equation}
where $\x$ and $\c$ are the ground truth speech and input features;
$\hat{S}$ and $S$ denote the spectral magnitude of output source excitation signal and the residual spectrogram;
$\psi$ and $N$ denote the function that transforms a spectral magnitude into the corresponding mel-spectrogram and the number of dimension of the mel-spectrogram, respectively.
This loss facilitates the output source excitation signal to have perceptually similar spectral characteristics as $S$ by using it as the target.

\subsubsection{Pitch dependent excitation generation}
$\text{F}_0$-driven architectures significantly improve the $\text{F}_0$ extrapolation capability of data-driven DNNs \cite{qpnet, qppwg, usfgan, hn-usfgan, nict1, nict2, nict3}.

USFGAN adopts pitch-dependent dilated convolution neural networks (PDCNNs) \cite{qpnet, qppwg} to improve $\text{F}_0$ controllability. 
Here, we define $F_\text{s}$, $f_t$, and $d$ as sampling frequency, $\text{F}_0$ value at time step $t$, and a constant dilation factor of a certain convolution neural network (CNN), respectively. 
In PDCNNs, the dilation size of the CNN layer dynamically changes for each $t$ depending on $f_t$.
The time-variant dilation size $d_t$ is calculated as follows:
\begin{equation} \label{}
    d_t = 
    \begin{cases}
        \lfloor E_t \rfloor \times d & \mbox{if} ~ E_t > 1 \\
        1 \times d & \mbox{else},
    \end{cases}
\end{equation}
where $E_t = F_\text{s} / (f_t \times a)$ is a proportional value to the period length modulated by a dense factor $a$. 
The constant dense factor controls the representable maximum frequency with the given sampling frequency $F_\text{s}$.

To further improve $\text{F}_0$ controllability, uSFGAN adopts sine waves as the inputs inspired by \cite{nsf}.
The sine waves directly provide the periodic information to the network leading to stable and fast learning and improved $\text{F}_0$ controllability.
Moreover, the effectiveness of the sine wave has been validated in several studies \cite{nsf, nsf_journal, periodnet, hn-pwg, nhv}.

\section{Source-Filter HiFi-GAN}
\label{sec:sifigan}
As shown in \figref{fig:sifigan}, SiFi-GAN consists of the source-network and filter-network connected in series. 
The source-network generates source excitation representation through an upsampling process and the representation is fed at each temporal resolution of the filter-network.

\subsection{Generator Architecture}

\subsubsection{Source excitation generation network}
We build the source-network with downsampling 1D CNNs, transposed 1D CNNs, and our designed quasi-periodic residual blocks (QP-ResBlocks).
The input sine wave, generated with the metrics of \cite{nsf}, goes through the stride CNNs to provide the resolution-matched periodic representations to the corresponding upsampling layer as proposed by \cite{nict1, nict2, nict3}.

\begin{figure}[tb]
\begin{center}
    \includegraphics[width=\columnwidth]{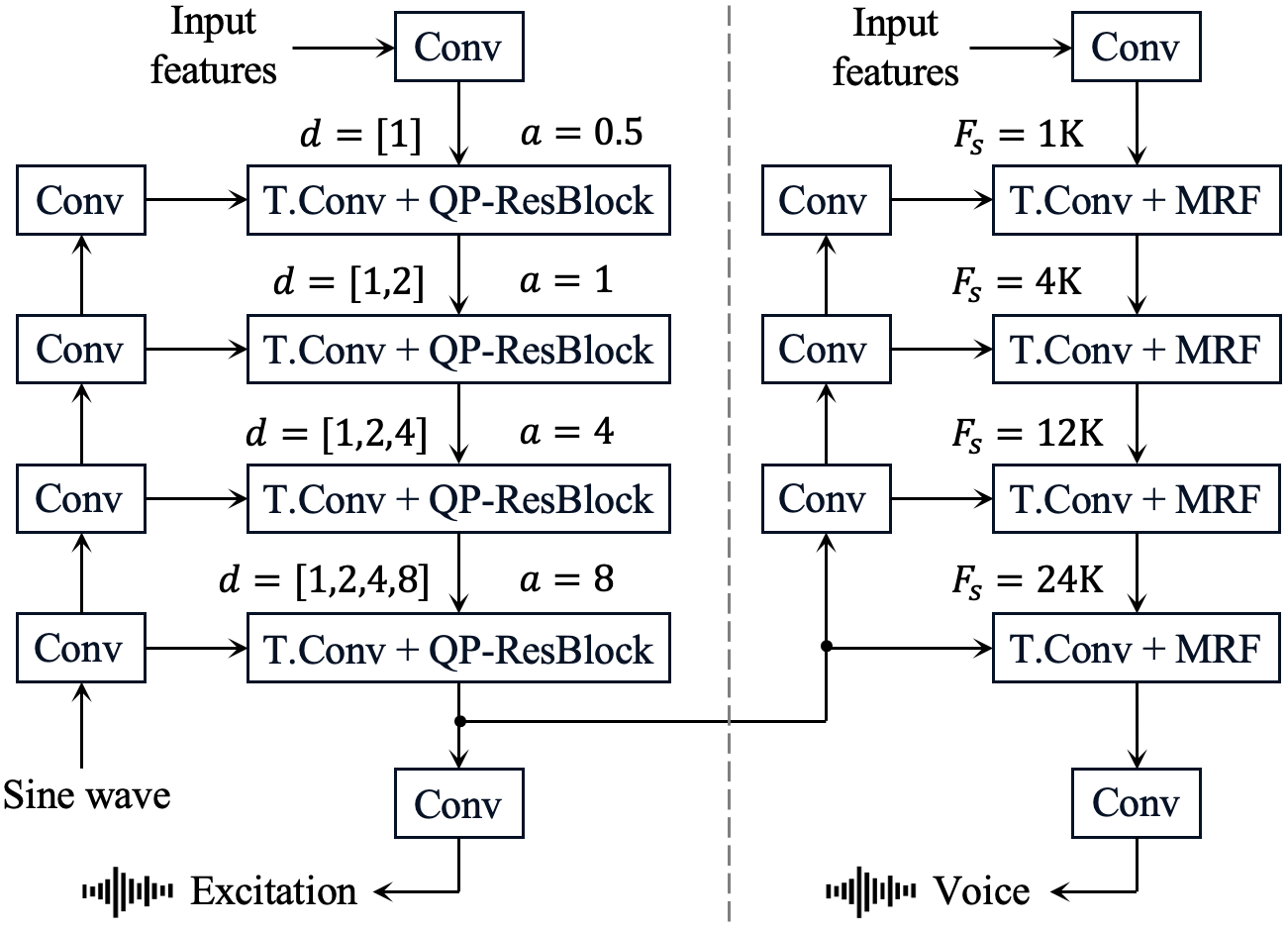}
    \vspace{-5mm}
    \caption{\fontsize{9.77}{11.5}\selectfont SiFi-GAN comprises the source-network (left) and filter-network (right). Conv, T.Conv, QP-ResBlock, and MRF denote 1D convolution, transposed 1D convolution, quasi-periodic residual block, and multi-receptive fusion \cite{hifigan}, respectively. $d$, $a$, and $F_{\text{s}}$ denote dilation sizes, the dense factor \cite{qpnet}, and the sampling rate in Hz, respectively.}
    \label{fig:sifigan}
\end{center}
\vspace{-4mm}
\end{figure}

The QP-ResBlock comprises of several iterations of leaky ReLU \cite{leakyrelu}, PDCNN \cite{qpnet, qppwg}, and 1D CNN with a residual connection after each iteration.
Each repetition comprises customized dilated sizes for each QP-ResBlock and the given dilation sizes define the number of repetitions.
We carefully set the dense factors \cite{qpnet} for each temporal resolution considering the theoretically producible maximum frequencies in PDCNNs.
In this paper, the dilation sizes and dense factors are set as shown in \figref{fig:sifigan}.
All kernel sizes in QP-ResBlocks are set to three.

Similar to uSFGAN \cite{usfgan, hn-usfgan}, the source excitation signal is regularized by further processing the outputs of the final QP-ResBlock through the leaky ReLU and 1D CNN.
The outputs of the final QP-ResBlock are also passed to the filter-network for the following voice generation.
Note that the pitch-dependent architectures are used only in the source-network.

\subsubsection{Resonance filtering network}
The filter-network is constructed as the original HiFi-GAN generator, except that the feature maps from the source-network are now added to the outputs of the transposed CNNs.
The feature maps are processed by additional downsampling CNNs configured as the sine embedding CNNs.
The final speech is obtained as the output of the filter-network through the same output layers of HiFi-GAN.

To compensate for the additional computation costs from the source-network, we configure the hyperparameters of MRFs in HiFi-GAN V1 to keep the synthesis speed fast.
First, the kernel sizes $\{ 3, 7, 11 \}$ are minimized to $\{ 3, 5, 7 \}$ to reduce the parameters and improve synthesis speed. 
We found that reducing the kernel sizes can greatly reduce the synthesis speed.
Furthermore, we excluded the additional CNNs following the dilated CNNs.

Notably, feeding the output of the final QP-ResBlock to the filter-network through downsampling CNNs is essential for the high-frequency reproduction in the output voice.
We show the details and the importance by an ablation study in Section \secref{sec:experimental evaluations}.

\subsection{Training Criteria}
The training procedure follows HiFi-GAN except that 
we adopt the regularization loss described in \mathref{math:reg-loss} and \cite{hn-usfgan} instead of the original feature matching loss in \cite{melgan, hifigan}.
The final training objective is formulated as follows:
\begin{flalign}
    \begin{split}
    \L_{G} &= \L_{g, \text{adv}} + \lambda_{\text{mel}}\L_{\text{mel}} + \lambda_{\text{reg}}\L_{\text{reg}}.
    \end{split}
\end{flalign}
We set the loss balancing hyperparameters $\lambda_{\text{mel}}$ and $\lambda_{\text{reg}}$ to 45.0 and 1.0 empirically.

\section{Experimental Evaluations}
\label{sec:experimental evaluations}
This section presents the performance of the proposed method.
Because $\text{F}_0$ controllability is more required in SVS, we evaluated singing voices generated with different $\text{F}_0$ scaling factors in the analysis-synthesis scenario. Audio samples and the code can be accessed from our demo site\footnote{\url{https://chomeyama.github.io/SiFiGAN-Demo}}.s

\subsection{Data preparation}
\label{sssec:data preparation}
We used Namine Ritsu's database \cite{namine_ritsu}, a Japanese singing voice dataset of 110 songs with a single female singer.
The total recording duration is about 4.35 hours. 
We split the songs using a ratio of 100/5/5 for training, validation, and test sets.
Also, each song was split into small clips based on the rest notes in the musical scores.
All data were downsampled from 44.1~kHz to 24~kHz and normalized to -26~dB.
The input acoustic features were extracted using WORLD \cite{world} with 5 ms frame shift. 
Specifically, we used one-dimensional continuous $\text{F}_0$ (c$\text{F}_0$) \cite{continuousF0}, one-dimensional voiced/unvoiced binary flags (v/uv), 40-dimensional mel-generalized cepstral coefficients (mgc), and three-dimensional band aperiodicity (bap). All sine waves were generated from c$\text{F}_0$ with the generation metric of \cite{nsf}.

\subsection{Model Details}
\label{sssec:model details}
We prepared three baseline models. 
1) $\textbf{WORLD}$ is a conventional source-filter vocoder \cite{world} with high $\text{F}_0$ controllability and reasonable voice quality.
2) $\textbf{hn-uSFGAN}$ denotes harmonic-plus-noise uSFGAN \cite{hn-usfgan}, which achieves high levels of voice quality and $\text{F}_0$ controllability. 
This model was conditioned on \{mgc, bap\}.
2) $\textbf{HiFi-GAN +~Sine}$ denotes HiFi-GAN \cite{hifigan} conditioned on \{c$\text{F}_0$, v/uv, mgc, bap\} and the sine embedding through downsampling CNNs \cite{nict1, nict2, nict3} as in SiFi-GAN.
This model is used to show the limitations of $\text{F}_0$ controllability with simple $\text{F}_0$ conditioning.
3) $\textbf{HiFi-GAN +~Sine~+~QP}$ denotes extended HiFi-GAN +~Sine by inserting QP-ResBlocks after each transposed CNN.
This model was used to show that just using the pitch-dependent mechanisms is insufficient to achieve high $\text{F}_0$ controllability.
Note that HiFi-GAN +~Sine and HiFi-GAN +~Sine +~QP have configurations roughly equivalent to \cite{nict1} and \cite{nict3}, respectively.
Because the vanilla HiFi-GAN conditioned on only the WORLD features cannot generate reasonable singing voices with transformed $\text{F}_0$, we omitted it for the following experiments.
However, the generated voices of the vanilla HiFi-GAN can be found on our demo page for reference.

An ablation model $\textbf{SiFi-GAN Direct}$ was also compared.
In this model, the source excitation representations from each QP-ResBlock are directly fed to filter-network at the corresponding temporal resolution without passing downsampling CNNs. 
Both the proposed $\textbf{SiFi-GAN}$ and SiFi-GAN Direct were conditioned on \{mgc, bap\}.

The upsampling rates of the upsampling layers were 5, 4, 3, and 2. 
The UnivNet multi-period and multi-resolution discriminators \cite{univnet} were adopted for all vocoders because of the effectiveness of the high-frequency component generations.
We trained all neural vocoders for 400~K steps with the same training settings of HiFi-GAN \cite{hifigan}.
The minibatch size was 16, and the minibatch length was 8400.
However, since the training of hn-uSFGAN takes longer than HiFi-GAN-based models, hn-uSFGAN was trained with minibatch size six.

\renewcommand{\arraystretch}{0.97}
\begin{table*}
\vspace{-2.4mm}
\centering
\caption{\fontsize{9.77}{11.5}\selectfont Results of objective and subjective evaluations. The MOS of the ground truth samples was $3.99 \pm 0.05$ ($1.0 \times \text{F}_0$).
Real time factors (RTFs) were computed with 108 clips (totally 878 s) on a single AMD EPYC 7542 or GeForce RTX 3090.}
\vspace{0.5mm}
\label{table:results}
\scalebox{1}{
\begin{tabularx}{2\columnwidth}{XYYYYYY}
\toprule
\small Metrics & \small WORLD & \small hn-uSFGAN & \small HiFi-GAN +~Sine & \small HiFi-GAN \ +~Sine~+~QP & \small \textbf{SiFi-GAN} \ (Proposed) & \small SiFi-GAN Direct \\ 
\midrule
\multicolumn{7}{c}{\cellcolor[HTML]{E6E6E6} \small Copy Synthesis ~ ($1.0 \times \text{F}_0$)} \\
\small V/UV [\%] $\downarrow$ & \small $4$ & \small $3$ & \small $2$ & \small $2$ & \small $2$ & \small $2$ \\
\small RMSE [Hz] $\downarrow$ & \small $0.09$ & \small $0.06$ & \small $0.06$ & \small $0.06$ & \small $0.06$ & \small $0.06$ \\
\small MOS $\uparrow$ & \small $2.98 \pm 0.07$ & \small $3.55 \pm 0.05$ & \small $3.86 \pm 0.05$ & \small $3.73 \pm 0.05$ & \small $\bm{3.90 \pm 0.05}$ & \small $3.78 \pm 0.05$ \\
\multicolumn{7}{c}{\cellcolor[HTML]{E6E6E6}\small $\text{F}_0$ transformation ~ ($0.5 \times \text{F}_0$)} \\
\small V/UV [\%] $\downarrow$ & \small $5$ & \small $3$ & \small $4$ & \small $4$ & \small $4$ & \small $5$ \\
\small RMSE [Hz] $\downarrow$ & \small $0.10$ & \small $0.09$ & \small $0.08$ & \small $0.09$ & \small $0.08$ & \small $0.10$ \\
\small MOS $\uparrow$ & \small $2.63 \pm 0.08$ & \small $2.97 \pm 0.07$ & \small $2.95 \pm 0.06$ & \small $3.24 \pm 0.06$ & \small $\bm{3.29 \pm 0.06}$ & \small $2.91 \pm 0.06$ \\
\multicolumn{7}{c}{\cellcolor[HTML]{E6E6E6}\small $\text{F}_0$ transformation ~ ($2.0 \times \text{F}_0$)} \\
\small V/UV [\%] $\downarrow$ & \small $7$ & \small $6$ & \small $19$ & \small $26$ & \small $10$ & \small $13$ \\
\small RMSE [Hz] $\downarrow$ & \small $0.11$ & \small $0.11$ & \small $0.22$ & \small $0.20$ & \small $0.13$ & \small $0.18$ \\
\small MOS $\uparrow$ & \small $2.74 \pm 0.08$ & \small $\bm{3.23 \pm 0.07}$ & \small $2.69 \pm 0.08$ & \small $2.61 \pm 0.09$ & \small $3.05 \pm 0.08$ & \small $2.87 \pm 0.07$ \\
\midrule
\small RTF (CPU) $\downarrow$ & -- & \small $3.97$ & \small $0.88$ & \small $1.13$ & \small $0.74$ & \small $0.73$ \\
\small RTF (GPU) $\downarrow$ & -- & \small $1.9 \times 10^{-1}$ & \small $3.0 \times 10^{-3}$ & \small $8.6 \times 10^{-3}$ & \small $6.2 \times 10^{-3}$ & \small $6.2 \times 10^{-3}$ \\
\# of Params $\downarrow$ & -- & \small $2.3 \times 10^6$ & \small $14.4 \times 10^6$ & \small $15.1 \times 10^6$ & \small $11.3 \times 10^6$ & \small $9.7 \times 10^6$ \\
\bottomrule
\end{tabularx}%
}
\vspace{-4mm}
\end{table*}
\renewcommand{\arraystretch}{1}


\subsection{Objective Evaluation}
\label{ssec:objective}
To investigate the synthesis efficiency, the real-time factors (RTFs) on a single CPU and GPU and the numbers of model parameters are reported.
To evaluate the $\text{F}_0$ controllability, root mean square error of log $\text{F}_0$ (RMSE) and voiced/unvoiced decision error rate ($\text{V/UV}$) are also reported.

The results shown in \tabref{table:results} can be summarized as follows:
1) SiFi-GAN achieves comparable RMSE and V/UV as WORLD and hn-uSFGAN, indicating its convincing $\text{F}_0$ controllability.
2) HiFi-GAN models show significant degradation for $\text{F}_0 \times 2.0$, implying a limitation of just adopting the pitch-dependent mechanism and the importance of source-filter modeling.
3) Compared to hn-uSFGAN, the upsampling mechanism of SiFi-GAN greatly improves the synthesis speed. 
4) Compared to the HiFi-GAN models, SiFi-GAN achieves better RTF with even fewer parameters.
Also, SiFi-GAN achieves faster synthesis speed on the CPU than the vanilla HiFi-GAN without sine embeddings nor QP-ResBlocks (RTF: 0.74 v.s. 0.84).

\begin{figure}[tb]
\begin{center}
    \includegraphics[width=\columnwidth]{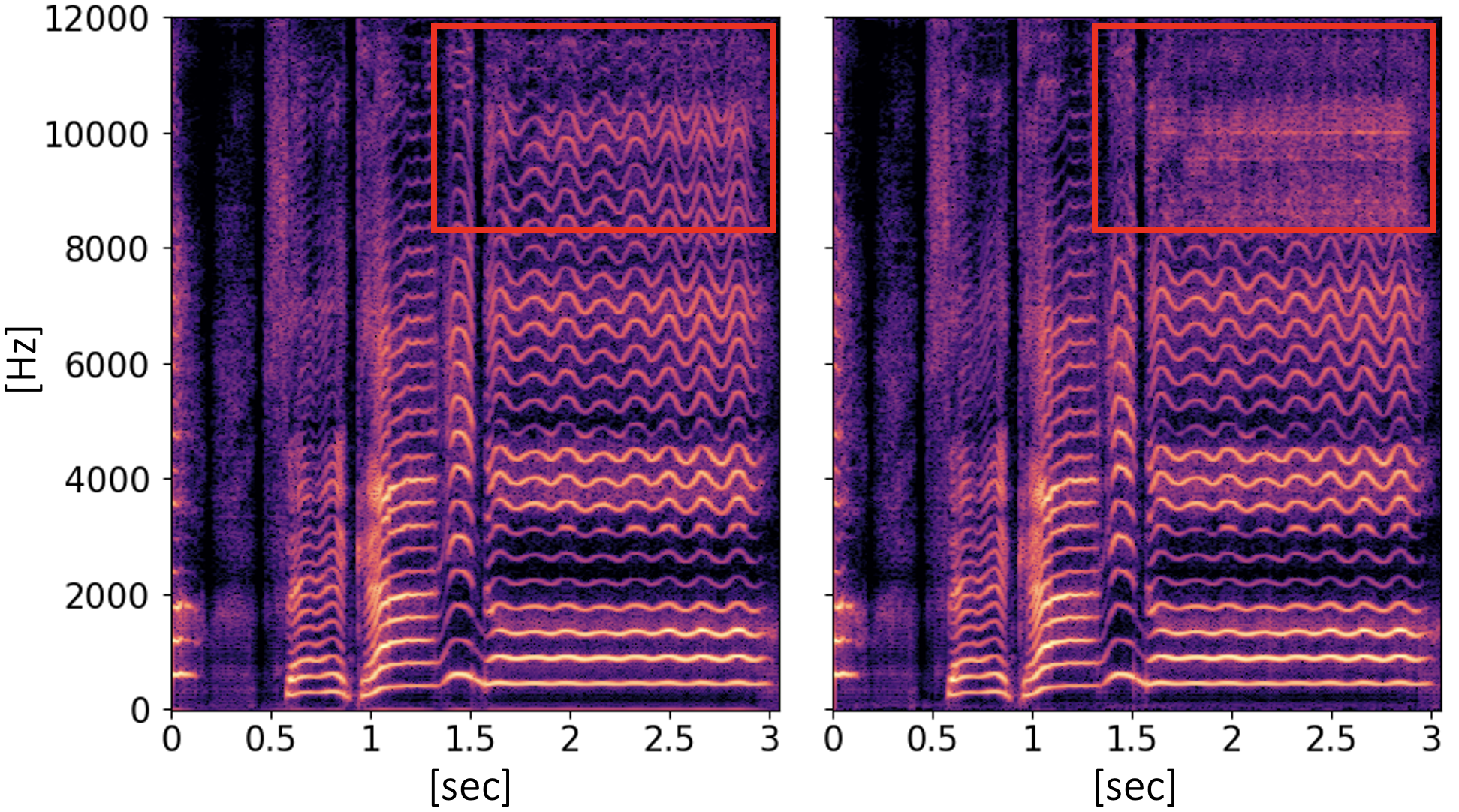}
    \vspace{-8mm}
    \caption{\fontsize{9.77}{11.5}\selectfont Spectrograms of output singing voices from SiFi-GAN (left) and SiFi-GAN Direct (right), respectively.}
    \label{fig:spec}
\end{center}
\vspace{-5mm}
\end{figure}

\subsection{Subjective Evaluation}
\label{ssec:subjective}
We conducted five scaled mean opinion scores (MOS) tests to evaluate the perceptual singing voice quality. 
Twenty Japanese subjects participated in the tests and each subject evaluated 12 samples for each method and each $\text{F}_0$ scaling factor.

The results shown in \tabref{table:results} can be summarized as follows:
1) SiFi-GAN outperforms WORLD and hn-uSFGAN significantly except for $\text{F}_0 \times 2.0$.
2) HiFi-GAN models show significant degradation for $\text{F}_0 \times 2.0$, while SiFi-GAN still achieves a high score.
3) SiFi-GAN Direct significantly degrades from SiFi-GAN for all $\text{F}_0$ scaling factors.

Specifically, we find that hn-uSFGAN usually suffers from buzzy noises while SiFi-GAN does not. 
The possible reason is that the handcraft sine wave inputs of hn-uSFGAN usually result in unnatural high-frequency components, but the learnable downsampling CNNs of SiFi-GAN provide more reasonable periodic information for each temporal resolution. 
Compared to the HiFi-GAN + Sine or + QP models, the better performance of SiFi-GAN indicates the effectiveness of the source-filter modeling. 

Moreover, as shown in \figref{fig:spec}, SiFi-GAN Direct easily fails to reconstruct the high-frequency components while SiFi-GAN attains clear high-frequency details. 
The result implies that the incomplete harmonic information of the intermediate excitation signals results in high-frequency modeling difficulties.
In conclusion, the upsampling source-filter architecture with the learnable downsampling CNNs of SiFi-GAN not only facilitates the synthesis efficiency but also provides more reasonable and tractable hierarchical harmonic information for corresponding temporal resolutions.

\section{Conclusion}
This paper introduced a high-fidelity neural vocoder SiFi-GAN with fast synthesis and high $\text{F}_0$ controllability. 
Since SiFi-GAN can generate raw waveforms in real-time, even with a single CPU, it can be applied to real-time applications.
Also, thanks to the fast training and inference speed, SiFi-GAN can be incorporated into end-to-end systems like TTS and SVS as an interpretable and $\text{F}_0$ controllable vocoder.
Although a significant gap exists between GT and generated voices, we believe the gap can be filled using input features with fewer estimation errors.

\noindent\textbf{Acknowledgement}: This work was supported in part by JST, CREST Grant Number JPMJCR19A3 and JSPS KAKENHI Grant Number 21H05054.

\newpage

\bibliographystyle{IEEEbib}
\setstretch{0.88}
\fontsize{9.3}{11.4}\selectfont
\bibliography{main}

\end{document}